\documentclass[11pt,psfig]{article}
\usepackage[american]{babel}
\usepackage[latin1]{inputenc}
\usepackage{amsfonts,amssymb,amstext}
\usepackage{amsmath,amsthm}
\usepackage{graphicx}
\usepackage{url}
\usepackage{hyperref}
\usepackage{xspace}
\usepackage{chngpage}
\usepackage{cmap}
\usepackage[usenames]{color}
\usepackage{epsfig}
\usepackage{color}
\usepackage{latexsym}
\usepackage{pstricks}
\usepackage{setspace}
\usepackage{algorithm}
\usepackage{algpseudocode}

\newtheorem{theorem}{Theorem}

\newtheorem{lemma}[theorem]{Lemma}

\newtheorem{definition}[theorem]{Definition}
\newtheorem{claim}[theorem]{Claim}

\def\Pr{\mathbb{P}}
\def\E{\mathbb{E}}

\def\beq{\begin{equation}}
\def\eeq{\end{equation}}
\def\beqn{\begin{eqnarray}}
\def\eeqn{\end{eqnarray}}

\title{Near-Optimal Radio Use For Wireless Network Synchronization}

\author{Milan~Bradonji\'c\thanks{Theoretical Division, and Center for Nonlinear Studies, Los Alamos National Laboratory, Los Alamos, NM 87545, USA. Email: milan@lanl.gov.
Current address: Mathematics of Networks and Communications, Bell Laboratories, Alcatel-Lucent, 600 Mountain Avenue, Murray Hill, New Jersey 07974, USA. Email: milan@research.bell-labs.com.},
Eddie Kohler\thanks{Computer Science Department, University of California Los Angeles, CA 90095, USA. Email: kohler@cs.ucla.edu.},
Rafail Ostrovsky\thanks{Computer Science Department and Department of Mathematics, University of California Los Angeles, CA 90095, USA. Email: rafail@cs.ucla.edu.
Research supported in part by NSF grants 0830803, 09165174, 1016540, 1065276, 1118126 and 1136174, US-Israel BSF grant 2008411, grants from OKAWA Foundation, IBM, Lockheed-Martin Corporation and the Defense Advanced Research Projects Agency through the U.S. Office of Naval Research under Contract  N00014-11-1-0392. The views expressed are those of  the author and do not reflect the official policy  or position of the Department of Defense or the U.S. ~Government.}
}

\date{}

\begin{document}
\maketitle

\begin{abstract}
In this paper, we consider the model of communication where
wireless devices can either switch their radios off to save energy
(and hence, can neither send nor receive messages), or switch
their radios on and engage in communication. The problem has been extensively studied in practice, in the setting such as deployment and clock synchronization of wireless sensor networks.

We distill a clean theoretical formulation of minimizing radio use and present near-optimal solutions. Our base model ignores issues of communication interference,
although we also extend the model to handle this requirement.  
We assume that nodes intend to communicate
periodically, or according to some time-based schedule.
Clearly, perfectly synchronized devices could switch their radios
on for exactly the minimum periods required by their joint
schedules.
The main challenge in the deployment of wireless networks is to \emph{synchronize} the devices'
schedules, given that their initial schedules may be offset
relative to one another (even if their clocks run at the same
speed).
In this paper we study how frequently the devices must switch on
their radios in order to both synchronize their clocks and
communicate. In this setting, we significantly improve previous results,  and show optimal use of the radio for two processors
and near-optimal use of the radio for synchronization of an
arbitrary number of processors.
In particular, for two processors we prove {\bf deterministic} matching $\Theta(\sqrt{d})$ upper and lower
bounds on the number of times the radio has to be on, where $d$ is
the discretized uncertainty period of the clock shift between the
two processors. (In contrast, all previous results for two processors are
randomized).
For $n=d^\beta$ processors (for any positive $\beta < 1$) we prove
$\Omega(d^{(1-\beta)/2})$ is the lower bound on the number of times
the radio has to be switched on (per processor), and show a nearly
matching (in terms of the radio use)
\~{O}$(d^{(1-\beta)/2})$ randomized
upper bound per processor.
For $\beta \geq 1$ our algorithm runs with at most $\mbox{{\em {\em poly-log}}}(d)$ radio invocations per processor.
Our bounds also hold in a radio-broadcast model where interference must be taken into account.

\noindent
\textbf{Keywords}: 
Wireless sensor network, clock synchronization, energy efficiency, algorithms, birthday paradox, probabilistic protocols.
\end{abstract}

\section{Introduction}
{\sc Motivation:} Radios are inherently power-hungry. As the power costs of
processing, memory, and other computing components drop, the
lifetime of a battery-operated wireless network deployment comes
to depend largely on how often a node's radio is left on.
System designers therefore try to power down those radios as much as
possible. This requires some form of \emph{synchronization},
since successful communication requires that the sending and
receiving nodes have their radios on at the same time.
Synchronization is relatively easy to achieve in a wired, powered,
and well-administered network, whose nodes can constantly listen
for periodic heartbeats from a well-known server.  In an ad hoc
wireless network or wireless sensor network deployment, the
problem becomes much more difficult.  Nodes may be far away from
any wired infrastructure; deployments are expected to run
and even to initialize themselves autonomously
(imagine sensors dropped over an area by plane); and environmental
factors make sensors prone to failure and clock drift.
Indeed there has been a lot of work in this area; see for example
\cite{BMT,S1,S2,bush05,ER,elson02,fan04P0PDIS,KORS,kopetz89,mills91,moscibroda06,mcglynn01,VPC,palchaudhuri04,polastre04versatile,sichitiu03,shnayder04simulating,S3,sivrikaya04,sundararamanBK05}.
Many distinct problems are considered in these papers, and it is beyond the scope of this paper to survey all these works, however most of these papers (among other issues) consider the following problem of radio-use consumption:

\bigskip
\noindent
{\sc Informal Problem description:}

Consider two (or more) processors that
can switch their radios on or off.  The processors' clocks are not
synchronized.
That is, when a processor wakes up, each clock
begins to count up from 0; however, processors may awake at
different times.  The maximum difference between the time when processors wake up is bounded by some positive integer parameter $d \in \mathbb{N}$.
If processors within radio range have their radios
on in the same step, they can hear each other and can synchronize
their clocks.  When a processor's radio is off, it saves energy,
but can neither receive nor transmit.
Initially, processors are awaken with clock shifts that differ by at
most $n$ time units. The objective for all the processors is to
synchronize their clocks while minimizing the use of radio (both
transmitting and receiving). We count the maximum number of times
any processor's radio has to be on in order to guarantee
synchronization. Indeed, as argued in many  papers referenced above, the total time duration during which the radio is on is one of the critical parameters of energy consumption, and operating the radio for considerable time is far costlier than switching radio off and switching it back on. We assume that all the processors that have
their radios on at the same time can communicate with each other.
The goal of all processors is to synchronize their clocks, i.e., to figure out how much to add to their offset so that all processors wake up at the same time.
(We also consider an extension that models radio
\emph{interference}, where if more then one processor is
broadcasting at the same time, all receiving processors that have their radio switched on hear only
noise.)

For multiple processors, we assume that all processors  know the
maximum clock offset $d$, otherwise the adversary can make the delay
unbounded. It is also assumed that all processors know the total number of processors $n$,
although, we also consider a more general setting where $d$ is known for all
processors, but $n$ is not.  In this setting, we relax the
problem, and instead of requiring synchronization of all $n$
processors, we instead require synchronization of an arbitrarily close to $1$
constant fraction of all processors.  In this relaxation of our
model, we require that the radio usage guarantee holds only for
those processors that eventually synchronize.

Furthermore, our model assumes that all processors are within radio
range of each other, so that the link graph is complete.
Our techniques can
be thought of as establishing synchronization within completely
connected single-hop regions.
Clearly, single-hop synchronization is necessary for multi-hop
synchronization.
Thus our central
concern in this paper is on establishing lower bounds and
constructing nearly optimal solutions for the single-hop case.

\bigskip
\noindent
{\sc Toward Formalizing the Abstract  Model:}

\noindent
{\sc A new model:}
To simplify our setting we wish to minimize both transmit and
receive cost (i.e., all the times when the radio must be ``on''
either transmitting or receiving). We discretize time to {\it time units} whose length
is equal to the smallest possible time that allows a processor to send a
message to or receive a message from another processor within radio
range. We normalize the cost of transmitting and receiving to one unit
of energy per one time unit. (In practice, transmission can be about
twice as expensive as receiving.  We can easily re-scale our
algorithms to accommodate this as well, but for clarity of
exposition we make these costs equal.) We ignore the energy consumption needed to power the radio on and to power it off, which is at most comparable but in many cases insignificant compared to the energy consumption of having the radio active.
This is the model considered, for example, in
\cite{mcglynn01,moscibroda06,S1,S2,S3,ZhengMobiHoc03,dutta08sensys,polastre04versatile}.

\noindent
{\sc Informal Model Description:}
For the purposes of analysis only, we assume that there is global
time (mapped to positive integers). All clocks can run at different
speeds, but we assume that clock drifts are bounded; i.e., there
exists a global constant $c$, such that for any two clocks their
relative speed ratio is bounded by $c$.
Now, we define as a time ``unit'' the number of steps of the slowest
clock, such that if two of the fastest processors' consecutive awake
times overlap by at least a half of their length according to global
time, then the number of steps of the slowest clock is sufficient
time for any two processors to communicate with each other.
This issue is elaborated in Section~\ref{sec:clockspeeds}.
In Section~\ref{sec:synchronization_algorithm}, we give the Synchronization Algorithm.
We now formalize the informal model description into the precise definition of our model.

\bigskip
\noindent
{\sc Our Formal Model and Problem Statement:}

Global time is
expressed as a positive integer. $n$ processors start at an
arbitrary global time between $1$ and $d$, where each processor
starts with a local ``clock'' counter set to 0. The parameter $d$
refers to the discretized uncertainty period, or equivalently, to
the possible maximal clock difference, i.e., to the maximal offset
between clocks; hence, we will use these terms interchangeably. Both
global time and each started processor's clock counter increments by
$1$ each time unit.
The global clock is for analysis only and is not accessible to any
of the processors, but an upper bound on $d$ is known to all
processors. Each processor algorithm is synchronous, and can
specify, at each time unit, if the processor is ``awake'' or
``sleeping.'' (The ``awake'' period is assumed to be sufficiently long to ensure that the energy consumption
 of powering the radio on and then shutting it off at each time unit is far less than the energy expenditure to operate the radio even for a single time unit). All processors that are awake at the same time unit
can communicate with each other. (Our interference model changes
this so that exactly two awake processors can communicate with each
other, but if three or more processors are simultaneously awake,
none of them can communicate.) The algorithm can specify what
information they exchange. The goal is for all $n$ processors to adjust
their local clocks to be equal to each other, at which point they
should all terminate. The protocol is correct if this happens either
always or if the protocol is randomized with probability of error
that is negligible. The objective is to minimize, per processor, the
total number of times its radio is awake.

We remark that the above model is sufficiently expressive to capture
a more general case where clocks at different nodes run at somewhat
different speeds, as long as the ratio of different speeds is
bounded by a constant, which is formally proven in Section~\ref{sec:clockspeeds}.

\bigskip
\noindent
{\sc Our Results:}

We develop algorithms for clock synchronization
in radio networks that minimize radio use, both with and without
modeling of interference.
In particular, our results are the following.
\begin{itemize}
\item[1.] For two processors we show a $\Omega(\sqrt{d})$ deterministic lower bound
and a matching deterministic $O(\sqrt{d})$ upper bound for the
number of time intervals a processor must switch its radio on to
obtain one-hop synchronization.
\item[2.] For arbitrary $n=d^\beta$
processors, we prove $\Omega\left(d^{\frac{1-\beta}{2}}\right)$ is
the lower bound on the number of time intervals the processor must
switch its radio for any deterministic protocol  and show a
nearly-matching (in terms of the number of times the radio is in
use) $O\big(d{\frac{1-\beta}{2}}  \mbox{{\em {\em
poly-log}}}(d)\big)$ randomized protocol, which fails to
synchronize with probability of failure exponentially (in $d$) close
to zero. Furthermore, our upper bound holds even if there is
interference, i.e., if more than one processor is broadcasting,
listening processors hear noise.
\item[3.] It is easy to see that processors cannot perform
synchronization if $d$ is unknown and unbounded, using a standard
evasive argument. However, if $d$ is known, we show that $8/9$
(or any other constant fraction) of the processors can synchronize
without knowing $n$, yet still using
$O(d^{\frac{1-\beta}{2}}  \mbox{{\em {\em poly-log}}}(d))$ radio
send/receive steps, with probability of failure exponentially close
to zero.
\end{itemize}

We stress that while the upper bound for two processors is simple,
the matching lower bound is nontrivial. This (with some additional
machinery) holds true for the multi-processor case as well.

\bigskip
\noindent
{\sc Comparison with Previous (Systems) Work:}
Tiny, inexpensive
embedded computers are now powerful enough to run complex software,
store significant amounts of information in stable memory, sense
wide varieties of environmental phenomena, and communicate with one
another over wireless channels.
Widespread deployments of such nodes promise to reveal previously
unobservable phenomena with significant scientific and
technological impact.
Energy is a fundamental roadblock to the long-lived deployment of
these nodes, however.
The size and weight of energy sources like batteries and solar
panels have not kept pace with comparable improvements to
processors, and
long-lived deployments must shepherd their energy resources
carefully.

Wireless radio communication is a particularly important energy
consumer.
Already, communication is expensive in terms of energy usage, and
this will only become worse in relative terms: the power cost of
radio communication is far higher than that of
computation.
In one example coming from sensor networks, a Mica2 sensor node's
CC1000 radio consumes almost as much current while listening for
messages as the node's CPU consumes in its most active state, and
transmitting a message consumes up to 2.5 times more current than
active CPU computation~\cite{shnayder04simulating}. In typical
wireless sensor networks, transmitting is about two times more
expensive than listening, and about 1.5 times more expensive than
receiving, but listening or transmitting is about 100 times more
expensive as keeping the CPU idle and the radio switched
off\footnote{Example consumption costs: CPU idle with clock
running and radio off (``standby mode''), 0.1--0.2~mA (milliamps);
CPU on and radio listening, 10~mA; CPU on and radio receiving,
15~mA; CPU on and radio transmitting, 20--25~mA. } (i.e., in a
``sleep'' state).

Network researchers have designed various techniques for
minimizing power consumption~\cite{S1,S2,S3}. For example,
Low-Power Listening~\cite{polastre04versatile} trades more
expensive transmission cost for lower listening cost. Every node
turns on its radio for listening for a short interval $\tau$ once
every interval $d > \tau$.  If the channel is quiet, the node returns to sleep for
another $d$; otherwise it receives whatever message is being
transmitted.  To transmit, a node sends a \emph{preamble} of at
least $n$ time units long before the actual message.  This ensures
that no matter how clocks are offset, any node within range will
hear some part of the preamble and stay awake for the message.
A longer $d$ means a lower relative receive cost (as $\tau/d$ is
smaller), but also longer preambles, and therefore higher
transmission cost.

A more efficient solution in terms of radio use was proposed by
 PalChaudhuri and Johnson~\cite{palchaudhuri04}, and further by
 Moscibroda, Von Rickenbach and Wattenhofer
~\cite{moscibroda06}. The idea is as follows.
Notice that in the proposal of \cite{polastre04versatile}, the proposal was for a
transmitting processor to broadcast continuously for $d$ time units,
while receiving processors switch their radios on once every $d$
time units to listen. Even for two processors, this implies that
total use of the radio is $d+1$ time units (i.e., it is linear in
$d$).  The observation of \cite{palchaudhuri04,moscibroda06} is that we can do substantially
better by using randomization: if both processors wake their radios
$O(\sqrt{d})$ time units at random (say both sending and receiving),
then by birthday paradox with constant probability they will be awake at the same time
and will be able to synchronize their clocks. As indicated before, we show instead a
deterministic solution to this problem, its practical importance, and a matching lower bound.

Our results strengthen and generalize previous works that appeared in the literature ~\cite{mcglynn01,ZhengMobiHoc03,tseng2003,dutta08sensys}. See further comparisons in the relevant sections.

\bigskip
\noindent
{\sc Comparison with Radio Broadcast:}

A node in a network, within a broadcast setup, is able to receive a message from a neighbor
only if it does not transmit, plus only one of its neighbors transmits at that time.
In the case when nodes are not able to detect collisions, there has been a body of literature that provides synchronization protocols among the nodes; for instance, for randomized broadcast algorithms see~\cite{BYGI,ABNLP}.
On the other hand, deterministic broadcast algorithms for the model without collision detection were proposed in~\cite{rchlebus02deterministic}:
an optimal linear-time broadcasting algorithm for symmetric graphs, as well as an algorithm for arbitrary graphs on $n$ nodes that works in time of order $O(n^{11/6})$.
The  improvements of these algorithms were given in~\cite{kowalski03broadcasting}: concretely, for undirected radio networks with diameter $D$, there was given a randomized
broadcast algorithm with the expected running time of order $O(D  \log (n /D) + \log ^2 n)$, while a deterministic broadcast algorithm had the expected running time of order $\Omega (n \log n / \log (n /D))$. Moreover, a faster algorithm for directed radio networks with running time of order $O(n \log n \log D )$ was provided in~\cite{kowalski03faster}. Finally, for the additional literature on other broadcast algorithms, we refer the reader to~\cite{EK,GM,Chiu-Yuen,KORS}.

The radio broadcast problem is a different from the problem addressed in our work.
We address the issue of near-optimal radio use and improve upon the previous results on clock synchronization algorithms~\cite{VPC,sichitiu03,ER,BMT}.
However, in order to avoid interference among the nodes, our solution can easily be combined with the radio broadcast problem.

We now give the high-level of constructions and proofs in our work.

\bigskip
\noindent
{\sc   High-level ideas of our constructions and proofs:}
\begin{itemize}
\item For the two processor upper bound, we prove that two carefully chosen affine functions will overlap
no matter what the initial shift is. The only technically delicate part is that the shift is over the reals, and thus the proof must take this into account.
\item For the two processor lower bound, we show that for any two strings with sufficiently low density (of $1$'s) there always exists a small shift such that none of the $1$'s overlap. This is done by a combinatorial counting argument.
\item For multiple processors, the idea of the lower bound is to extend the previous combinatorial argument, while for the upper bound, the idea is to establish a ``connected" graph of pairwise processor synchronization, and then show that this graph is an expander. The next idea is that instead of running global synchronization, we can repeat the same partial synchronization a logarithmic number of times (using the same randomness) to yield a communication graph which is an expander. We then use standard synchronization protocol over this ``virtual" expander to reach global synchronization.
\item For handling interference, we observe that standard ``back-off" protocols~\cite{aldous1987,cali2000} can be combined with previous machinery to achieve  non-interference, costing only a poly-logarithmic multiplicative term.
\item For the protocol that does not need to know $n$ (recall that $n$ is the total number of processors within radio-reach), we first observe that if $n>d$, by setting $n=d$ our protocol already achieves synchronization with near-optimal radio use. The technical challenge is thus to handle the case where $n < d$ but the value of $d$ is unknown to the protocol.  Our first observation is to show that processors can overestimate $n$, in which case the amount of energy needed is much smaller (per processor) than for smaller $n$, and then ``check'' if the synchronized component of nodes has reached the current estimate on $n$.  If it did not, than our current estimate of $n$ could be reduced (by a constant factor) by all the processors.
To assure that estimates are lowered by all the processors at about the same time, we divide the protocol into ``epochs'' which are big enough not to overlap even with a maximal clock offset (of $d$). Summing, the energy consumption is essentially dominated by the smallest estimate of $n$,
which is within a constant factor of the correct value of $n$, and all processors that detect it stop running subsequent (more expensive) ``epochs''.
\end{itemize}

\section{Mathematical Preliminaries}
In this section, we first state  the Chernoff bound and some property of the floor function that will be widely used in our exposition. Then we claim and prove Lemma~\ref{two-color-birthday-lemma} and
Lemma~\ref{ex-two-color-birthady-lemma}.
\begin{enumerate}
\item[(i)] Chernoff Bound: Let $X_1,X_2,\dots,X_L$ be independent Bernoulli random variables with
$\Pr[X_i=1]=p_i$. Denote $X = \sum_{i=1}^L X_i$ and $\mu = \E[X]$.
For any $\delta \in (0,1)$ the following is satisfied:
\beq
\nonumber
\Pr\left[X < (1-\delta)\mu\right] < \left(\frac{e^{-\delta}}{(1-\delta)^{1-
\delta}}\right)^\mu, \quad \Pr\left[X > (1+\delta)\mu\right] <
\left(\frac{e^{\delta}}{(1+\delta)^{1+\delta}}\right)^\mu.
\eeq
\item[(ii)] For any $x \in \mathbb{R}$, $x = \lfloor x \rfloor +
\{x\}$, where $\{x\} \in [0,1)$ is the fractional part, and
$\lfloor x \rfloor \in \mathbb{Z}$ is the floor function, the
following is satisfied:
\[
\lfloor x \rfloor -1 \leq x -1 < \lfloor
x \rfloor \leq x < \lfloor x \rfloor + 1 \leq x+1.
\]
\end{enumerate}

\begin{lemma}[Two-Color Birthday Problem]\label{two-color-birthday-lemma}
For any absolute constant $C > \sqrt{1 - \ln 0.1} \approx 1.8173$ and any
positive $s, t \in (0,1)$, where $s + t = 1$, the following holds. Suppose
$r=C L^s$ identical red balls and $b=C L^t$ identical blue balls
are thrown independently and uniformly at random into $L$ bins. Then, for sufficiently big $L$,
the probability that there is a bin containing both red and blue balls is
$\geq 0.8$.
\end{lemma}

\begin{proof}
We first note that since all balls are thrown independently and uniformly at random,
it follows that throwing all $r+b$ balls together uniformly at random, is equivalent to the scenario of
first throwing $r$ red balls, then throwing $b$ blue balls. Thus, we first throw $r$ red balls, and count the number of unoccupied bins. Let $Z$ be a random variable, denoting the number of empty bins, after throwing $r=CL^s$ balls into $L$ bins, u.a.r. The expectation of $Z$ is given by
\begin{displaymath}
\E[Z ] = L \left(1 - \frac{1}{L} \right)^r.
\end{displaymath}
By the occupancy bound, Theorem 1 in~\cite{kamath95tail}, for any $\theta>0$, the tail of $Z$ satisfies
\begin{displaymath}
\Pr[|Z - \E[Z ] | \geq \theta \E[Z] ] \leq 2 \exp \Big(- \frac{\theta^2 \E[Z]^2 (L-1/2)}{L^2 - \E[Z]^2} \Big).
\end{displaymath}
Now, let us throw $b$ blue balls into $L$ bins, u.a.r. Some of these $L$ bins  have been  previously occupied with red balls. Let us denote the event $\mathcal H$ that exists at least one bin with both red and blue balls. (The goal is to show $\Pr[\mathcal H]>0.8$ for the given assumptions.) Given $Z=z$, the probability that one blue ball does not hit a bin with a red ball is $z/L$. Then it follows $\Pr[\mathcal H | Z=z] = 1-(z/L)^b$, and furthermore,
\begin{displaymath}
\Pr[\mathcal H | Z \leq (1+\theta)\E[Z]] \geq 1-((1+\theta)\E[Z]/L)^b,
\end{displaymath}
where we will appropriately choose $\theta = o(1)$, later. We now use the total probability formula and the bound on $Z$, in order to obtain a lower bound on the probability of $\mathcal H$:
\begin{eqnarray}
\nonumber \Pr [ \mathcal H] &=& \Pr \Big[\mathcal H | Z \leq (1+\theta)\E[Z] \Big] \Pr \Big[Z \leq (1+\theta)\E[Z] \Big] \\
\nonumber  && + \Pr \Big[\mathcal H | Z > (1+\theta)\E[Z] \Big] \Pr \Big[Z > (1+\theta)\E[Z]\Big]\\
\nonumber &>& \Pr \Big[\mathcal H | Z \leq (1+\theta)\E[Z] \Big] \Pr \Big[Z \leq (1+\theta)\E[Z] \Big]\\
\nonumber &\geq& \Big( 1-((1+\theta)\E[Z]/L)^b \Big) \Big(1 -2 \exp \Big(- \frac{\theta^2 \E[Z]^2 (L-1/2)}{L^2 - \E[Z]^2} \Big) \Big) \\
\nonumber &=& \Big( 1-(1+\theta)^b(1-1/L)^{rb} \Big) \Big(1 -2 \exp \Big(- \frac{\theta^2 \E[Z]^2 (L-1/2)}{L^2 - \E[Z]^2} \Big) \Big).
\end{eqnarray}
Let us choose $\theta = 1/b = 1/(CL^t)$. The goal is to obtain a lower bound on $\Pr[\mathcal H]$, sufficiently close to 1, so we discuss the following terms.
First,
\begin{eqnarray}
\nonumber 1 - (1+\theta)^b \left(1-\frac{1}{L}\right)^{rb}  &=& 1- \left(1+\frac{1}{b}\right)^b \left(1-\frac{1}{L}\right)^{C^2 L^{s+t}} \\
\nonumber &=& 1 - \left(1+\frac{1}{CL^t} \right)^{CL^t} \left(1-\frac{1}{L} \right)^{C^2 L}\\
\nonumber &\geq& 1 - e^{1 - C^2},
\end{eqnarray}
where we have used $(1+1/L)^L \leq e$ and $(1-1/L)^L \leq 1/e$, for every $L$. (Furthermore, the sequences $(1+1/L)^L$ and $(1-1/L)^L$ are both increasing, with the limits $e$ and $1/e$, respectively.)
Second, let us consider the term $\theta^2 \E[Z]^2 (L-1/2)/(L^2 - \E[Z]^2)$. We have

\begin{eqnarray}
\nonumber \frac{\theta^2 \E[Z]^2 (L-1/2)}{L^2 - \E[Z]^2} &\geq& \frac{\theta^2 L /2}{(L/\E[Z])^2-1}\\ 
\nonumber &= & \frac{\theta^2 L}{2} \frac{(1-1/L)^{2r}}{1-(1-1/L)^{2r}} \\
\nonumber &\geq & \frac{\theta^2 L}{2} \frac{(1-1/L)^{2r}}{2r/L} \\
\nonumber &= & \frac{L}{2 b^2} \frac{(1-1/L)^{2r}}{2r/L} \\
\nonumber &=& \frac{r}{4C^4} (1-1/L)^{2r} \,,
\end{eqnarray}
where we used: (i) the expression for $\E[Z]$, (ii) Bernoulli's inequality $(1 + x)^r \geq 1 + rx$ for  $x>-1$ and $r \geq 1$,  (iii) $\theta = 1/b$, (iv) $rb = C^2 L$
by definitions for $r$ and $b$.

In order to find a lower bound on the last expression, let us consider the logarithm value of it:
\begin{eqnarray}
\nonumber \ln \frac{r}{4C^4} (1-1/L)^{2r} &=& -\ln(4C^4) + \ln r + 2r\ln(1-1/L) \\
\nonumber &=& -\ln(4C^4) + \ln r - 2r \left( \frac{1}{L} + O\left(\frac{1}{L} \right) \right) \\
\nonumber & =& -\ln(4C^4) + \ln C + s \ln L - \frac{2}{L^{1-s}} + O\left(\frac{1}{L^{1-s}}\right) \\
\nonumber &\geq& \frac{s}{2} \ln n \,,
\end{eqnarray}
for sufficiently large $L$. Now it follows
\begin{eqnarray}
\nonumber 1 - 2\exp\Big(-\frac{\theta^2 \E[Z]^2 (L-1/2)}{L^2 - \E[Z]^2}\Big) &\geq& 1 - \exp(-  \frac{r}{4C^4} (1-1/L)^{2r} ) \\
&\geq& 1 - \exp( - L^{s/2}).
\end{eqnarray}
Thus, for sufficiently large $L$, we obtain the lower bound on the probability of $\mathcal H$
\begin{eqnarray}
\label{eq:Hbnd}
\nonumber \Pr[\mathcal H] &\geq& (1 - e^{1 - C^2})(1 - 2\exp( - L^{s/2}))\\
& = & 1 - e^{1 - C^2} - 2\exp( - L^{s/2}) + 2e^{1 - C^2} \exp( - L^{s/2}).
\end{eqnarray}
For a given $\epsilon \in (0,1)$, if both $\epsilon /2  \geq e^{1 - C^2}$ and $\epsilon /4 \geq \exp( - L^{s/2})$, then Eq.~(\ref{eq:Hbnd}) implies $\Pr[\mathcal H] \geq 1 - \epsilon$. These two conditions are equivalent to $c \geq \sqrt{1 - \ln (\epsilon/2)}$, and $n \geq ( -\ln (\epsilon/4 ) )^{2/s}$.
Finally, we obtain $\nonumber \Pr[ \mathcal H] = 1 - \epsilon$. Specifically, let $C > \sqrt{1 - \ln 0.1} \approx 1.8173$, that is $1 - e^{1 - C^2} >0.8$. Then for sufficiently big $L$, it follows $\Pr[\mathcal H] \geq 0.8$, which completes the proof.
\end{proof}

\textbf{Note.} \emph{In the next sections, without loss of
generality we round any real number $x$ to an integer, by using the ceiling function
$\lceil x \rceil$, e.g., we treat $L^\alpha$, $L/C^2$ as integers.}

\begin{lemma}[Exclusive Two-Color Birthday Problem]\label{ex-two-color-birthady-lemma}
For any absolute constant $C \leq 5$ and for any positive $s,t \in
(0,1)$, where $s + t = 1$, and $L$ sufficiently large, the
following holds. Consider $r=C L^s$ identical red balls and $b=C L^t$
identical blue balls thrown independently and uniformly at random into
$L$ bins. The probability that there is a bin with exactly one red
and one blue ball is greater than $3/4$.
\end{lemma}
\begin{proof}
Let $R \subseteq [L]$ be the set of bins occupied by exactly one
red ball; similarly let $B \subseteq [n]$ be the set of bins
occupied by exactly one blue ball. We estimate the cardinalities
of these two sets, and then show that their intersection is
nonempty with probability at least $3/4$, for sufficiently
large $L$.

Let $X_i$ be a random variable such that $X_i=1$ when there is
exactly one red ball in the $i$th bin, otherwise $X_i=0$. We
have $\Pr[X_i=1] = { r \choose 1} \frac{1}{L}(1-\frac{1}{L})^{r-1}$. Furthermore, for $X=\sum_{i=1}^L X_i$ we have
$\E[X] =r(1- \frac{1}{L})^{r-1}$. Since $r=C L^s$ and $0<s<1$, that
is $r/L = 1/(C L^s) = o(1)$, it follows $\E[X] =C L^s (1-o(1))$. For
some constant $\delta \in (0,1)$, using Chernoff Bound, it follows
$\Pr[ X \leq (1-\delta)\E[X]] \leq \exp \left(-\E[X] \delta^2/2\right) \to 0$
as $L \to +\infty$, since $\E[X] =C L^s (1-o(1)) \to +\infty$. Analogously,
we define $Y_i$ and $Y$ for blue balls. Since $0<t<1$
it follows $\E[Y] =b(1- \frac{1}{L})^{b-1}$. We have $\Pr[ Y \leq
(1-\delta)\E[Y]] \leq \exp \left(-\E[Y] \delta^2/2\right) \to 0$ as $L \to
+\infty$, since $\E[Y] =C L^t (1-o(1)) \to +\infty$.

Let $x = (1-\delta)\E[X]$ and $y = (1-\delta)\E[Y]$ be the expected
cardinalities of the sets $R$ and $B$, respectively (the balls are thrown uniformly
and independently into the bins). Let $\mathcal{T}$ be the event that
there exists a bin with exactly one red ball and one blue ball. Then it follows
\begin{eqnarray}
\nonumber \Pr[\mathcal{T}] &\geq&  \Pr[\mathcal{T} | X >
(1-\delta)\E[X], Y > (1-\delta)\E[Y]] \\
\nonumber && \times \Pr[X > (1-\delta)\E[X], Y > (1-\delta)\E[Y]]\\
\nonumber &\geq& \Pr[\mathcal{T} | X = x, Y = y] \Pr[X >
(1-\delta)\E[X], Y > (1-\delta)\E[Y]]\\
\nonumber &=& \Pr[\mathcal{T} | X = x, Y = y] \Pr[X >
(1-\delta)\E[X]] \Pr[Y > (1-\delta)\E[Y]]\\
\nonumber &\geq& \Pr[\mathcal{T} | X = x, Y = y] (1 - e^{-\E[X]
\delta^2/2})(1 - e^{-\E[Y] \delta^2/2}).
\end{eqnarray}

Given the cardinality of the set $R$, any $|R|$-combination from
$[L]$ is equally probable. Similarly it follows for the set $B$. Hence
\begin{eqnarray} \Pr[\mathcal{T}^c |
X = x, Y = y] &=& \nonumber \frac{{L \choose x}{L - x \choose
y}}{{L \choose x}{L \choose y}} = \frac{(L-x)!(n-y)!}{L!(L-x-y)!}.
\end{eqnarray}

Taking the logarithm of the last expression, and for sufficiently large $L$ since the values $x, y, x+y$ are
$o(L)$, it follows

\begin{eqnarray}
\nonumber
\ln \Pr[\mathcal{T}^c | X = x, Y = y] &=&
\sum_{k=L-x-y+1}^{L-y}\ln k - \sum_{k=L-x+1}^{L}\ln k \\
\nonumber
&=& -\sum_{k=y}^{x+y-1} \ln (1 + \frac{y}{L-k}) \\
\nonumber
&\geq& -\sum_{k=y}^{x+y-1} \frac{y}{L-k} \\
\nonumber
&\geq& - y \int_{L-x-y}^{L-y} \frac{\textrm{d}x}{x} \\
\nonumber
&=& - y \ln \left( 1 + \frac{x}{L-x-y} \right)\\
\nonumber
&=& - \frac{xy}{L} \left(1 + o(1)\right) \\
\nonumber
&=& -\frac{C^2 (1-\delta)^2 L^{s+t}}{L}(1+o(1))\\
\nonumber
&=& -C^2 (1-\delta)^2 (1+o(1)) \,.
\end{eqnarray}

Finally, for $\delta = 1/10$ and $C \leq 5$ we have $\exp \left(-C^2(1-\delta)^2 \right)
\geq 0.7788$, that is, $\Pr[\mathcal{T}] \geq 3/4$ for
sufficiently large $L$.
\end{proof}

\section{Lower Bounds}
\label{sec:lower-bounds}
The problem of asynchronous wakeup, i.e., low-power asynchronous neighbors discovery, has already been known in the literature~\cite{mcglynn01,ZhengMobiHoc03,tseng2003,dutta08sensys}.
Its goal is to design an optimal wake-up schedule, i.e., to minimize the radio use for both transmitting and receiving. The techniques used, e.g., in the previously cited papers, vary from the birthday paradox in~\cite{mcglynn01}, block-design in ~\cite{ZhengMobiHoc03},
the quorum based protocol in \cite{tseng2003} to an adaption of Chinese remainder theorem~\cite{knuth} in~\cite{dutta08sensys}.
In our work, we first generalized the birthday paradox, by obtaining the Two Color Birthday Problem (see Lemma~\ref{two-color-birthday-lemma}). Next, we build the tools for our main analysis on the upper and lower bounds on the optimal radio use for wireless network synchronization.
In particular, we start with Lemma~\ref{lm:2ncs}, which is a stronger combinatorial
bound compared to~\cite{ZhengMobiHoc03}, and then generalize results in Lemma~\ref{general-non-colliding}.

Recall that $d$ is the maximum offset between processor starting
times and $n = d^\beta$ is the number of processors. Assume that
each processor runs for some time $L$. Its radio schedule can then
be represented as a bit string of length $L$, where the $i$th bit
is 1 if and only if the processor turned its radio on during that
time unit. We first consider the two-processor case.
Recall that in our model maximal assumed offset is at most $d$.
If we take $2$ bit strings corresponding to the two processors, the initial clock
offset corresponds to a {\em shift} of one string against the
other by at most $d$ positions. Note that if we set
$L\geq 4d$, the maximal shift is at most
$d \leq L/4$.

To prove our lower bound, we need to prove the following: for any
two $L$-bit strings with at most $\sqrt{L}/C$ ones in each string
(for some constant $C > 1 /\sqrt 2$), there always exists a shift
$< L/4$ of one string against another such that none of the ones
after the shift in the first string align with any of the ones in
the second string. In this case, we say that the strings do not
{\em overlap}. W.l.o.g.\, we make both strings (before the shift)
identical. To see that this does not limit the generality, we note
that if the two strings are not identical, we can make a new
string by taking their bitwise OR, which we call the
\emph{union} of strings. If the distinct strings overlap at a
given offset, then the union string will overlap with itself
at the same offset.

\begin{lemma}[Two Non-Colliding Strings]
\label{lm:2ncs}
For any absolute constant $C \geq 1/\sqrt{2}$, and for every
$L$-bit string with $\leq \sqrt L/C$ ones, there is at
least one shift within $L/(2C^2)$ such that the string and its
shifted copy do not overlap.
\end{lemma}
We want to prove a general lower bound for multiple strings.
The high-level approach of our proof is as follows. We pick one
string, and then upper bound the total number of ones possible in
the union of all the remaining (potentially shifted) strings.  If
we can prove that assuming the density of all the strings is
sufficiently small, and there always exists a shift of the first
string that does not overlap with the union of all the remaining
strings, the proof is completed. The union string will simply be a new string
with a higher density.

\begin{lemma}[General Two Non-Colliding Strings with Different Densities]
\label{general-non-colliding}
Let $s,t>0$ such that $s+t<1$, and let $C>1$. For two $L$-bit strings such that the number of ones in the first string is $a = L^s/C$,
and the number of ones in the second string is $b = L^t/C$, there is a shift up to $L/C^2+1$ such that the first string and the shifted second string do not overlap.
\end{lemma}
\begin{proof}
Let the positions of ones in the first string  be $P = \{ p_1, p_2, \dots, p_a\} \subseteq [L]$, and the positions of ones in the second string be $Q = \{ q_1, q_2,
\dots, q_b\} \subseteq [L]$. Let us consider the set of differences $\mathcal{I} = \{ p - q \mid p \in P, q \in Q \}$. The cardinality of $\mathcal{I}$ satisfies $|\mathcal{I}| \leq  |P| |Q|
= ab = L^{s + t}/C^2 \leq L/C^2 < L$.

Similarly to the proof of Lemma $\ref{lm:2ncs}$,
let us choose $i \in \{0,1,2, \dots, \lceil L/C^2 \rceil + 1\}$ such
that $i \notin \mathcal I$. That integer $i$ exists since
$|\mathcal I | \leq L/C^2$. Then $P$ and $Q + i =
\{q + i \mid q \in Q \}$ do not intersect, since by construction $p \neq q +
i$ for any $p \in P$ and any $q \in Q$.
\end{proof}

Here, w.l.o.g.\, we considered only the ``left'' shift. If we needed both
left and right shifts, then we would have an additional factor of $2$.
Using Lemma $\ref{general-non-colliding}$, the lower bounds
immediately follow.
\begin{theorem}
There exists an absolute constant $C>1$, such that for any
$d^\beta$ strings of length $L$ with at most
$d^{(1-\beta)/2}$ ones in each string, there always exists a
set of shifts for each string by at most $L/4$ such that no
string's ones overlap any of the ones in all the other strings.
\end{theorem}

\begin{proof}
Set $\alpha = (1-\beta)/2$. Add strings sequentially and for
each find a shift that does not overlap with (the union of) all
the shifted previous strings. Lemma $\ref{general-non-colliding}$
applies since the smaller string has density $d^\alpha$, and the
union of all the previous strings has density of at most $d^\beta d^\alpha$.
This density is at most $d$, since the sum of the exponents $\alpha + \beta = (1+\beta)/2$ is at most $1$, which proves the theorem.
\end{proof}

\section{Matching  Upper Bound for Two Processors}

We now show the upper bound. That is, we give the deterministic algorithm for
two devices.  In particular, for any initial offset of at most $d$,
we show a schedule where two processors ``meet'' with probability equal to one inside a ``time-window" of
length $W=2d + 4\sqrt d +2$.

\begin{theorem}\label{thm-comb}
For any $d$, there exists a string of length $W=2d + 4\sqrt d +2$
with at most $4\sqrt d +4$ ones such that this string will overlap
itself for all shifts from $1$ to $d$.
\end{theorem}
\begin{proof}
Let us define the string $S$ of length $W$, which has ones at the following positions (from the perspective of its local clock): Set the bits at positions $(i \sqrt d + i)$ and $(i \sqrt d)$ to 1, for $i \in \{1,\dots, \lfloor 2\sqrt d +2 \rfloor\}$; Set the remaining bits to $0$.

For the analysis, we consider the ``global'' clock. Furthermore, we consider two strings $S_1$ and $S_2$, being the shifted versions of the string $S$, and shifted by $a_0$ and $b_0$, respectively. (Both $a_0$ and $b_0$ are $\leq d$, by the conditions of Theorem $\ref{thm-comb}$.)
Since the string $S$ is deterministically defined, we know the exact appearances of ones in the strings $S_1$ and $S_2$.
Thus, from the global clock point of view, in the strings $S_1$ and $S_2$, respectively, ones appear during the following time intervals:
\begin{displaymath}
[\lfloor a_{i_1} \rfloor, \lfloor a_{i_1} \rfloor +1], [\lfloor a_{i_2}
\rfloor, \lfloor a_{i_2} \rfloor +1] \textrm{ and } [\lfloor b_{i_1} \rfloor, \lfloor b_{i_1} \rfloor +1], [\lfloor b_{i_2} \rfloor, \lfloor b_{i_2} \rfloor +1],
\end{displaymath}
where the values $a_{i_1}, a_{i_2}, b_{i_1},
b_{i_2}$ are given by: $a_{i,1} =  a_0 + i \sqrt d + i, a_{i,2} = a_0 + i \sqrt d, b_{i,1} = b_0 + i \sqrt d + i, b_{i,2} = b_0 + i \sqrt d$,
for $i \in \{1,\dots, \lfloor 2\sqrt d +2 \rfloor\}$. The
initial values of strings are: $a_{0,1}=a_{0,2}=a_0$ and $b_{0,1}=b_{0,2}=b_0$.
We next show that there exist integers $i, j \in \{0,1, \dots, \lfloor
2\sqrt d +2 \rfloor\}$, such that for some $s_1,s_2 \in \{1,2\}$ the following is satisfied
\begin{equation}
\label{eq:delta}
\delta = |a_{i,s_1}-b_{j,s_2}|<1.
\end{equation}
Although the schedule we propose may look simple, in general $\sqrt d$ is not an integer, thus we have to perform the precise analysis below.
%
%
So, we now explicitly construct $i,j$ such that Eq.~$(\ref{eq:delta})$ is satisfied, that is,
$a_{i,s_1}=b_{j,s_2} \pm \delta$ for some fractional part $\delta
\in [0,1)$ and for some $s_1,s_2 \in \{1,2\}$. Let us call the absolute difference $\Delta$ between the initial values of the strings $S_1$ and $S_2$, that is, $\Delta =|a_0-b_0|$.
From $a_0, b_0 \in \{0,1,\dots,n\}$ it follows $\Delta \leq d$, and we consider the following three cases.

\noindent
\textbf{Case 1: $a_0 - b_0 = 0$}.
It follows $a_0=b_0$ and Eq.~$(\ref{eq:delta})$ is satisfied for $i=j=0$.

\noindent
\textbf{Case 2: $1 \leq a_0-b_0 \leq d$}.
Let us define $q = \lfloor \Delta / \sqrt d \rfloor$.
Then it follows
\beqn
\nonumber
q &=& \left\lfloor \frac{\Delta}{\sqrt d} \right\rfloor \leq \frac{\Delta}{\sqrt d} \leq \frac{d}{\sqrt d} = \sqrt d\,, \\
\nonumber
\left\lfloor \frac{\Delta}{\sqrt d} \right\rfloor &=& \frac{\Delta}{\sqrt d} < \left\lfloor \frac{\Delta}{\sqrt d} \right\rfloor +1 \,.
\eeqn
Moreover
$q \leq \Delta / \sqrt d < q +1$ and $q  \sqrt d \leq \Delta < q \sqrt d + \sqrt d$.
We now give the exact $i$ and $j$ where the strings meet. Consider
$i = \lfloor (q+1)\sqrt d \rfloor - \Delta$ and $j = i + q+1$, then it follows
\begin{eqnarray}
\nonumber i &=& \lfloor (q+1)\sqrt d \rfloor - \Delta \leq
(q+1)\sqrt d - \Delta = q \sqrt d - \Delta + \sqrt d \leq \sqrt d,\\
\nonumber j &=& q+1+i \leq \sqrt d +1 + \sqrt d = 2 \sqrt d +1.
\end{eqnarray}
Since $\Delta$ is an integer it follows $\Delta
\leq \lfloor (q+1)\sqrt d \rfloor \Rightarrow i\geq 0 \Rightarrow
j>0$. By substituting the values for $i$ and $j$ we obtain $\delta =
|a_{i,1} - b_{j,2}| = \{(q+1) \sqrt d \} \in [0,1)$.

\noindent
\textbf{Case 3: $-d \leq a_0-b_0 \leq -1$}.
For the previously defined $i,j,\Delta$ it follows $|a_{j,2} - b_{i,1}| \in [0,1)$.

Finally, for any $a_0, b_0 \in \{0,1,\dots,d\}$ there are $i,j$ such that the shifted strings meet. Since $\max_{i,j} \{ a_i, b_j
\} \leq d + (2\sqrt d +2)\sqrt d+ (2\sqrt d +2) = 3d + 4\sqrt d +2$, and subtracting $d$, which is the length of the strings, it follows that
the strings meet with probability equal to $1$ inside the time-window of length $W=2d + 4\sqrt d +2 $, which proves the theorem.
\end{proof}

\begin{figure}[htb]
\linethickness{0.75mm}
\setlength{\unitlength}{0.5mm}
\begin{minipage}[h]{0.45\linewidth}
\begin{picture}(50,40)
\put(0,50){\line(1,0){98}}
\put(6,50){\line(0,-1){25}}
\put(7,50){\line(0,-1){25}}
\put(12,50){\line(0,-1){25}}
\put(18,50){\line(0,-1){25}}
\put(21,50){\line(0,-1){25}}
\put(28,50){\line(0,-1){25}}
\put(30,50){\line(0,-1){25}}
\put(35,50){\line(0,-1){25}}
\put(36,50){\line(0,-1){25}}
\put(42,50){\line(0,-1){25}}
\put(48,50){\line(0,-1){25}}
\put(49,50){\line(0,-1){25}}
\put(54,50){\line(0,-1){25}}
\put(56,50){\line(0,-1){25}}
\put(60,50){\line(0,-1){25}}
\put(63,50){\line(0,-1){25}}
\put(66,50){\line(0,-1){25}}
\put(70,50){\line(0,-1){25}}
\put(72,50){\line(0,-1){25}}
\put(77,50){\line(0,-1){25}}
\put(78,50){\line(0,-1){25}}
\put(84,50){\line(0,-1){25}}
\put(91,50){\line(0,-1){25}}
\put(98,50){\line(0,-1){25}}
\put(0,0){\line(1,0){98}}
\put(6,0){\line(0,1){25}}
\put(7,0){\line(0,1){25}}
\put(12,0){\line(0,1){25}}
\put(18,0){\line(0,1){25}}
\put(21,0){\line(0,1){25}}
\put(28,0){\line(0,1){25}}
\put(30,0){\line(0,1){25}}
\put(35,0){\line(0,1){25}}
\put(36,0){\line(0,1){25}}
\put(42,0){\line(0,1){25}}
\put(48,0){\line(0,1){25}}
\put(49,0){\line(0,1){25}}
\put(54,0){\line(0,1){25}}
\put(56,0){\line(0,1){25}}
\put(60,0){\line(0,1){25}}
\put(63,0){\line(0,1){25}}
\put(66,0){\line(0,1){25}}
\put(70,0){\line(0,1){25}}
\put(72,0){\line(0,1){25}}
\put(77,0){\line(0,1){25}}
\put(78,0){\line(0,1){25}}
\put(84,0){\line(0,1){25}}
\put(91,0){\line(0,1){25}}
\put(98,0){\line(0,1){25}}
\end{picture}
\end{minipage}%
\begin{minipage}[h]{0.45\linewidth}
\begin{picture}(50,40)
\put(10,50){\line(1,0){98}}
\put(16,50){\line(0,-1){25}}
\put(17,50){\line(0,-1){25}}
\put(22,50){\line(0,-1){25}}
\put(28,50){\line(0,-1){25}}
\put(31,50){\line(0,-1){25}}
\put(38,50){\line(0,-1){25}}
\put(40,50){\line(0,-1){25}}
\put(45,50){\line(0,-1){25}}
\put(46,50){\line(0,-1){25}}
\put(52,50){\line(0,-1){25}}
\put(58,50){\line(0,-1){25}}
\put(59,50){\line(0,-1){25}}
\put(64,50){\line(0,-1){25}}
\put(66,50){\line(0,-1){25}}
\put(70,50){\line(0,-1){25}}
\put(73,50){\line(0,-1){25}}
\put(76,50){\line(0,-1){25}}
\put(80,50){\line(0,-1){25}}
\put(82,50){\line(0,-1){25}}
\put(87,50){\line(0,-1){25}}
\put(88,50){\line(0,-1){25}}
\put(94,50){\line(0,-1){25}}
\put(101,50){\line(0,-1){25}}
\put(108,50){\line(0,-1){25}}
\put(0,0){\line(1,0){98}}
\put(6,0){\line(0,1){25}}
\put(7,0){\line(0,1){25}}
\put(12,0){\line(0,1){25}}
\put(18,0){\line(0,1){25}}
\put(21,0){\line(0,1){25}}
\put(28,0){\line(0,1){25}}
\put(30,0){\line(0,1){25}}
\put(35,0){\line(0,1){25}}
\put(36,0){\line(0,1){25}}
\put(42,0){\line(0,1){25}}
\put(48,0){\line(0,1){25}}
\put(49,0){\line(0,1){25}}
\put(54,0){\line(0,1){25}}
\put(56,0){\line(0,1){25}}
\put(60,0){\line(0,1){25}}
\put(63,0){\line(0,1){25}}
\put(66,0){\line(0,1){25}}
\put(70,0){\line(0,1){25}}
\put(72,0){\line(0,1){25}}
\put(77,0){\line(0,1){25}}
\put(78,0){\line(0,1){25}}
\put(84,0){\line(0,1){25}}
\put(91,0){\line(0,1){25}}
\put(98,0){\line(0,1){25}}
\end{picture}
\end{minipage}
\caption{Two strings will have at least one 1 aligned, for any right shift with non-empty overlap. For instance, for two strings of length $L=96$ with $d=36$, we present two right shifts of length $0$ and $10$, respectively.}
\label{fig-strings}
\end{figure}
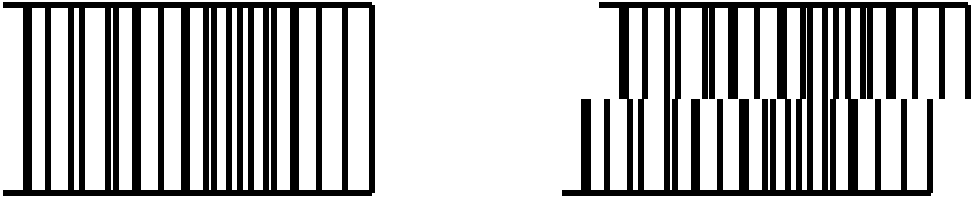

As an example of a ``good solution'', we present two identical strings of length $L = 2d + 4 \sqrt{d} + 2 = 98$, for $d=36$. For any right shift of length $1, \dots, d$, the two strings ``meet''. This is schematically given in Fig.~\ref{fig-strings}, for $d=36$ and two right shifts of length $0$ and $10$, respectively.

\bigskip

We remark that the bound proved in this section is in fact more general than the subsequent independent work of~\cite{dutta08sensys}, which appeared  after our report~\cite{bradonjic-2008-arxiv}. Note that our bound holds for all values of $d$ and the two strings could be made identical by doubling the cost.

\section{Upper Bound for $m$ Processors}

In this setting we have $n=d^\beta$ processors (and as before the
maximum shift is at most $d$). We first state our theorem:

\begin{theorem}
There exists a randomized protocol for $d^\beta$ processors (which fails with probability at most $1/2^{O(d)}$) such that:
(i) if $\beta < 1$ the protocol is using at most
$O\big(d^{\frac{1-\beta}{2}}\mbox{{\em poly-log}}(d)\big)$
radio steps per processor, and
(ii) if $\beta \geq 1$ using at most
$O\left(\mbox{{\em poly-log}}(d)\right)$ radio steps per processor.
Furthermore, the same bounds hold for the synchronization in the
radio communication model, where a processor can hear a message if
one (and strictly one) message is broadcast.
\end{theorem}

In the sequel, a high-level outline of the construction of our algorithm for $\beta \in [0,1)$. For the case of $\beta \geq 1$ we only need Step 4 and Step 5, see below.

\subsection{Outline of the Main Algorithm}
\label{sec:main.algo}
\begin{itemize}
\item[{\bf Step 1.}]
We let each processor run for $L=4 d$ steps,
waking up during this time $O(d^{\frac{1-\beta}{2}})$
times uniformly at random. It is important to point out that each
processor uses independent randomness. We view it as an $n$-row and
$L$-column $(L\geq W+d)$ matrix $A$ (taking into account all the
shifts), where $W = 2d+4 \sqrt d +2$ is defined in Theorem $\ref{thm-comb}$.
Fix any row of this matrix (say the first one). We say that this row
``meets'' some other row, if 1 in the first row also appears
(after the shifts) in some other row. If this happens, the first
processor can ``communicate'' with another processor.  We show
that for a fixed row, this happens with a constant probability.

\item[{\bf Step 2.}] Each processor repeats Step 1 (using
independent randomness) $O(\log n)$ times. Here, we show that a
fixed row has at least $O(\log n)$ connections to other rows (not
necessarily distinct) with probability greater than $1-1/\textrm{poly}(n)$.

\item[{\bf Step 3.}] From Step 2, we conclude that the first row
meets at least a constant number of {\em distinct} other rows with probability
greater than $1 - 1/(2n)$.

\item[{\bf Step 4.}] We use the union bound to conclude that {\em
every} row meets at least a constant number of distinct other rows with probability
greater than $1/2$. If we repeat this process a logarithmic number of times,
we show that we obtain an expander graph with overwhelming probability (for the definition of an expander see~\cite{motwani95raghavan}). Thus,
considering every row (i.e., every processor) as a node, this represents
a random graph with degree of at least a constant number for each node, which is an
expander with high probability.

\item[{\bf Step 5.}] During the synchronization period, a
particular processor will synchronize with some other processor,
without collision, by attempting to communicate whenever it has a 1
in its row. (In the case of interference, the processor can
communicate if only one other processor is up at this column,
which we can achieve as well, using standard ``back-off" protocols~\cite{aldous1987,cali2000}, costing only a poly-logarithmic multiplicative term.)

\item[{\bf Step 6.}] The processors can now communicate along the
edges of the formed expander (which has logarithmic diameter) as follows. The
main insight that we prove below is that if processors repeat {\em the same random choices} of
Step 1 through Step 5, the communication pattern of  the expander graph is preserved.
Hence, the structure developed in Step 2 can be reused to establish a
logarithmic-diameter (in $n$) spanning tree and synchronize nodes
with poly-logarithmic overhead (using known machinery over this
``virtual" graph). We show in Section~\ref{sec:synchronization_algorithm}, by using standard
methods, that communicating over the implicit expander graph to
synchronize all nodes can be done in $D+2$ steps, where $D$ is the
diameter of the expander.
\end{itemize}

\subsection{Proof of Correctness of the Main Algorithm}
\bigskip
{\bf Analysis of Step 1.} We generate a random matrix $A$ as follows.
\begin{definition}[Generation of the Random Matrix $A$]\label{gen-rand-matr-a}
For each row of the $d^\beta \times L$ random matrix $A$,
independently of the content of other rows, we uniformly and
independently generate $C L^\alpha$
integers $t_1, t_2, \dots, t_{C L^\alpha}
\in \{1,2,\dots,L\}$, where $\alpha,\beta \in [0,1]$. Each $t_i$ corresponds
to one energy unit for the unit time $t_i$, of that row.
\end{definition}
Note that $t_i$'s are not necessarily different, and the sum of each row is $C L^\alpha$.

\begin{lemma}
\label{lm:one-row-meeting}
Let $A$ be the matrix (given by Definition $\ref{gen-rand-matr-a}$), such that $2\alpha+\beta=1$. Let us consider one
particular row from $A$. That row ``meets" with some other row with probability $>0.8$.
\end{lemma}

\begin{proof}
In Section~\ref{sec:lower-bounds}, we used $L = 4 d$. Each row in the matrix $A$ has $\hat{C} L^\alpha = \hat C d^\alpha$ ones, where we denote
the constant $\hat C  = 4^\alpha C$.

Let us consider a particular row of the matrix $A$, w.l.o.g.\ let
it be the first row. The first row has $\hat C d^\alpha$ ones, which we call
the blue balls. 
Let all of the remaining $d^\beta -1$ rows be collapsed into one row, which we
call the ``collapsed row", and each entry of  this collapsed row represents one of $d$ bins.
The number of ones in the collapsed row is
\begin{displaymath}
(d^\beta -1) \hat C d^\alpha = d^{\alpha + \beta} \hat C \left(1 - \frac{1}{d^\beta}\right).
\end{displaymath}
The ones in the collapsed row, we call the red balls.

Since the positions of the balls in each row are generated independently and uniformly at random,
also row by row independently, it follows that the balls in the collapsed row are
generated independently and uniformly at random. That is, the
process of  throwing the first $\hat C d^\alpha$ red balls into
``collapsed bins", then the second $\hat C d^\alpha$ (red) balls, and
so on $d^\beta-1$ times, is equivalent to throwing $d^{a+\beta} \hat C (1 -d^{-\beta})$ red balls, all at once, into $d$ bins, independently and uniformly at random.
Now, by Lemma \ref{two-color-birthday-lemma} (Two-Color Birthday Lemma) it follows that for
$\alpha+(\alpha+\beta) = 2\alpha+\beta=1$, the particular (the first) row
``meets'' with some other row with probability $>0.8$.
\end{proof}

\bigskip
 {\bf Analysis of Step 2.}
We repeat $O(\log n)$ times the procedure ``Generation of the random matrix $A$."
Such constructed matrices concatenated to each other form the matrix
$\hat{A}$ of the dimension $n^\beta \times \Theta (L \log n)$.

For the sake of brevity, let us call $N := n-1$. Let us generate $l = K \log N$ random matrices
$A^{(1)}, \dots, A^{(l)}$, where $K$ is a constant to be
determined later in the analysis of Step 4. That is, let us repeat $l$ times the procedure
``Generation of the random matrix $A$."
We prove that each row in
the matrix $A$ has $0.4 K\log M$ ``meetings" with probability
$>1/2$. Again, w.l.o.g.\ let us consider the first row. Let
$X_i$ be a random zero-one variable, indicating that the first row ``meets" with some
other row in the matrix $A^{(i)}$, for $i=1,\dots,l$.
The variables $X_i$ are independent Bernoulli trials, since matrices $A^{(i)}$ are generated independently.
By Lemma $\ref{lm:one-row-meeting}$ it follows $\Pr[X_i=1]>0.8$.
Let $T = \sum_{i=1}^{l} X_i$ be the number of these ``meetings", then
\begin{displaymath}
\E[T]=\sum_{i=1}^{l} \E[X_i] > 0.8l = 0.8 K\log N.
\end{displaymath}
Applying the Chernoff Bound on $T$ it follows,
\begin{displaymath}
\Pr[T \leq (1-\delta)\E[T]] \leq \exp(-\E[T] \delta^2 /2) < e^{-0.4 l
\delta^2 }.
\end{displaymath}
Taking $\delta = 1/2$ we have the following bound on $T$
\begin{displaymath}
\Pr[T \leq 0.4 K\log N] \leq \Pr[T \leq \E[T] / 2] < e^{-0.1 K \log N} = N^{-0.1K}.
\end{displaymath}

{\bf Analysis of Step 3.} Here, we prove that
each row in the matrix $A$ ``meets" with at least a constant number of different
rows with probability $> 1/2$. We specify this constant later.
(Furthermore, the meetings are
chosen independently with replacement.) Again, w.l.o.g.\ we
consider the first row.
The number of ``meetings" is
$T > 0.4 K\log N$ with probability at least $1- N^{-0.1K}$.

We see this experiment as throwing $T$ balls, one by one,
independently, into $N$ bins. Let $Y_i$ be a binary random
variable, such that, $Y_i=1$ if and only if the number of already occupied
bins is increased by one, with the $i$th thrown ball, otherwise
$Y_i=0$. Note that the variables $Y_i$ are not independent, so that we cannot apply the Chernoff bound. We have $\Pr[Y_i=1] \geq 1-(i-1)/N > 1-T/N$, for
$i=1,\dots, T$. Let us denote $\Pr[Y_i=1]=p_i$ and $\Pr[Y_i=0]=q_i$, where $p_i \leq 1$, and $q_i<T/N$ for every $i$.
Let the value $Y=\sum_{i=1}^T Y_i$ be the number of occupied bins after throwing all $T$ balls.

The random variable $Y$ corresponds to the number of different rows that one particular (the first) row in $A$ ``meets". We show that the probability that $Y$ is less than some constant is upper bounded by $N^{-2}$.
W.l.o.g.\, we choose this constant to be $10$, otherwise the proof below applies for any other positive constant. We have
\begin{eqnarray}
\nonumber \Pr[Y \leq 10] &=&  \sum_{k=0}^{10} \sum_{I: I \subset
[T], |I|=k} \prod_{i \in I} p_i \prod_{j \in [T] \setminus I} q_j
\leq \sum_{k=0}^{10} \sum_{I: I \subset [T], |I|=k} (T/N)^{T-k}\\
\nonumber &=& \sum_{k=0}^{10} {T \choose k} (T/N)^{T-k} <
\sum_{k=0}^{10} \frac{T^T}{k! N^{T-k}} < \frac{T^T}{N^{T-10}}\sum_{k=0}^{10} \frac{1}{k!} \\
\nonumber &<& e T^T/N^{T-10}.
\end{eqnarray}
For $T > 0.4 K \log N >12$ it follows
\begin{displaymath}
e T^T/N^{T-10} = \exp(1+ T \log T - (T-10)\log N) < \exp(-2 \log N)=N^{-2},
\end{displaymath}
which completes the proof of Step 3.

\bigskip
{\bf Analysis of Step 4.} Taking $K$ such that $0.1K>1$
and $K \log N>30$, by the union bound, from the analysis of the previous steps, it follows that
\begin{eqnarray}
\nonumber \Pr[\textrm{the first row has at least $10$ different ``meetings"}] &>& 1-(N^{-0.1K}+N^{-2})\\
\nonumber &>& 1 - \frac{1}{2(N+1)} \\
\nonumber &=&1-\frac{1}{2n}.
\end{eqnarray}
Finally, by the union bound applied over all of $n$ rows, it follows that
every row has at least $10$ different ``meetings" with probability
$>1/2$, that is,
\begin{displaymath}
\Pr[\textrm{every row has at least 10 different ``meetings"}]>1/2\,.
\end{displaymath}

The matrix $\hat{A}$ uniquely defines a random graph $\hat{G}$,
for which we show that has the minimum degree at least $10$,
with probability $> 1/2$. Here we define the undirected graph $G=(V,E)$ that corresponds to the random matrix $A$.
\begin{definition}
For a graph $G=(V,E)$, the set of nodes $V=\{v_1, v_2, \dots, v_n\}$ corresponds to the
set of rows of the matrix $A$ (i.e., to the set of devices). For $1
\leq i < j \leq n$ the edge $(i,j) \in E$, if and only if there is
a column $t$ in the matrix $A$ such that $A_{i,t} \neq 0$ and
$A_{j,t} \neq 0$, for $1 \leq t \leq d$.
\end{definition}

Let $\hat G =(V, \hat E)$ be the graph with the set of nodes $V$
and the set of edges $\hat E = \cup_{i=1}^{l} E^{(i)}$ obtained as the union of edges
corresponding to the matrices $A^{(i)}$, for $i=1,\dots,l$.
We have proven that with probability $>1/2$ every vertex in the
graph $\hat G$ has degree $\geq 10$. Finally, we can repeat this entire process another
{\em {\em poly-log}}($d$) times to guarantee a success probability
exponentially close to 1.

\bigskip
 {\bf Analysis of Step 5.}
A fixed row in a
matrix $\hat A$ will meet with some other row, without collision,
with probability $\geq 1 - 0.4^{\Theta(\log n)} = 1 - n^{-\Theta(1)}$. In case of interference, use standard back-off protocol analysis, with a multiplicative overhead of $O(\log^2 n)$.

\bigskip
{\bf Analysis of Step 6.} We recall that if we have a random graph with node-degree at least a constant, then we can use the
following theorem:
\begin{theorem}[Bollobas, de la Vega \cite{bollobasdelavega}]
A random $\ell$-regular graph on $n$ nodes has diameter $(\log n +
\log \log n)/\log(\ell-1) + c$, for same small constant $c<10$. This
is the best possible since any $\ell$-regular graph has diameter at
least $\log n / \log(\ell-1)$.
\end{theorem}

In the graph $\hat G$, the degree of each node is at least $\ell$ (w.l.o.g.\ we have specifically chosen $\ell=10$).
Furthermore, by our construction, the edges are independent.
It follows that the $diam(\hat G)$ is at most the
diameter of a random $\ell$-regular subgraph. That is, $diam(\hat
G) = O(\log n)$ with high probability.

\begin{definition}
We say that an $n \times d$ zero-one matrix
$B=(b_{i,j})$ is associated with an undirected graph $G = (V,E)$
if and only if: the set of nodes is $V=[n]$, and
between two nodes $i \neq j$ there is an edge $(i,j) \in E$ if and only of if there is a column $t \in [d]$ in the
matrix $B$ such that $b_{i,t}=b_{j,t}=1$, and
$b_{k,t}=0$ for all $k \in [d]\setminus \{i,j\}$.
We also say that the graph $G$ is associated with the matrix $B$.
\end{definition}
Every processor $i \in [n]$ generates $O(d \log^2 n)$ random
variables
\beq
\nonumber
C^{i}_1,C^{i}_2,\dots, C^{i}_{O(d \log^2 n)},
\eeq
repeating $O(\log^2 n)$ times the procedure \emph{(Generation of
the Random Matrix $A$)} (see Definition \ref{gen-rand-matr-a}). That is, $i$ randomly
generates a string of length $d$, with exactly $d^\alpha =
d^{(1-\beta)/2}$ ones, while the rest of the entries are zeros. That string is
mapped onto $C^{i}_1,C^{i}_2,\dots, C^{i}_{d}$. Then $i$,
independent of the previous outcomes, repeats \emph{(Generation
of the Random Matrix $A$)} for the next $d$ variables
$C^{i}_{n+1},C^{i}_{n+2},\dots, C^{i}_{2d}$, and so on; totally
repeating $O(\log^2 n)$ times the procedure \emph{(Generation of the Random Matrix $A$)}.

We define the zero-one matrix $\widetilde{A}$, such that
$\widetilde{A}_{i,j} = C^{i}_{j}$, for $1 \leq i \leq n$, $1 \leq
j \leq O(d \log^2 n)$; that is, the $i$th row corresponds to the
coin outcomes of the $i$th processor. According to the way
the random matrix $\widetilde{A}$ of size $n \times O(d \log^2 n)$ is created,
$\widetilde{A}$ can be divided into $O(d \log n)$ blocks of the matrices $\hat{A}$'s, each of the
size $n \times O(d \log n)$. Finally, each of these $\hat{A}$'s
matrices, can be subdivided into $O(d \log n)$ blocks of the
matrices $A$'s, each of the size $n \times d$.
For matrices $A, \hat{A}$, $\widetilde{A}$, let
the associated graphs be $G, \hat{G}$, $\widetilde{G}$,
respectively.
In the Analysis of Step 2, we have proven that a particular row in
$\hat{A}$ has at least $10$ meetings with probability $>1/(2n)$. That is,
every node in $\hat{G}$, has a degree of at least $10$ with probability
$>1/2$. Then it follows that every row in $\widetilde{A}$ has at
least $10$ meetings with probability close to $1-n^{-\Theta(1)}$,
i.e., every node in $\widetilde{G}$ has a degree of at least 10 with
probability $1- n^{-\Theta(1)}$.
Finally, let us define $CommGraph$.
\begin{definition}
Let $CommGraph$ be graph whose incident matrix of size $n \times O(D d \log^2 n)$ is obtained by concatenating $D$
{\bf identical} copies of the matrix $\widetilde{A}$ of size $n \times O(d \log^2 n)$.
\end{definition}
We will later call this operation a {\em concatenation in time}. The term is motivated by the fact that the number of columns in $A, \hat{A}, \widetilde{A}$ represents the number of time steps over which the radio devices communicate (the radio devices may be on or off during this time period).

The processors will be able to communicate over {\em CommGraph} in time and
synchronize their clocks' drifts. The synchronization scheme and the
proof are given by the following Synchronization Algorithm.

\section{Synchronization Algorithm}
\label{sec:synchronization_algorithm}
Every processor $i$ has its own identification $ID_i$, which is a
random number. Let the number of random bits, representing an
$ID_i$, be much larger then $\log n$. Then it follows that all
$ID_i$'s are different, with probability arbitrarily close to one.
Furthermore, we will use the terms node and processor interchangeably.

Every node knows $n$, so that it can compute $D = O(\log n)$.
Also, every node $i$ keeps the following variables: $Max(i)$, a
set of neighbors $Neighbors(i)$, and $RootTime(i)$. Besides
$RootTime(i)$, a node keeps its own local-clock time,
$OwnTime(i)$. We now explain the variables.

At the beginning, the initialization for any node $i$ is the
following. Every node assumes that it has the maximum $ID$,
that is $Max(i)=ID_i$, at the beginning. The set of neighbors $Neighbors(i)$ is the
set of neighbors $i$ in the $CommGraph$,
$\widetilde{G}$, and the set of neighbors $Neighbors(i)$ is known
to $i$. The $RootTime(i)$ is set to the node's current time,
$RootTime(i)=OwnTime(i)$. Every node $i$ performs the
Synchronization Algorithm, defined below.

\begin{algorithm}[h!]
\caption{Synchronization Algorithm}
\label{alg:synchalgo}
\begin{algorithmic}[1]

\State
Send its $Max(i)$ and $RootTime(i)$ to the set of its neighbors
$Neighbors(i)$.
\vspace*{1mm}
\State
$\textrm{cnt} = 0$
\vspace*{1mm}
\Repeat
\vspace*{1mm}
\If {the node $i$ hears from a node $j$ and $Max(j) > Max(i)$}
\vspace*{1mm}
\State
Set $Max(i) := Max(j)$
\vspace*{1mm}
\State
Set the new $RootTime(i):=RootTime(j)+ \Delta_{tr}$
\vspace*{1mm}
\State
Propagate new $Max(i)$ and $RootTime(i)$ to
$Neighbors(i) \setminus \{j\}$
\vspace*{1mm}
\EndIf
\vspace*{1mm}
\State
$\textrm{cnt} := \textrm{cnt} +1$
\vspace*{1mm}
\Until {$\textrm{cnt} \leq D$}
\vspace*{1mm}
\State
Set own clock to the time of the node with the maximal $ID$, i.e., $OwnTime(i) = RootTime(i)$.
\vspace*{1mm}
\end{algorithmic}
\end{algorithm}

We now explain Algorithm~\ref{alg:synchalgo}. Line 1: The node $i$ transmits
$Max(i)$ and $RootTime(i)$ to the set of its neighbors. Line 3: Then the node $i$ listens $D$ times.
Line 4: If
the node $i$ hears from a node $j$ and $Max(j) > Max(i)$ then,
Line 5: We propagate $Max(ID)$;
Line 6: $i$ must update the time, $RootTime(i)$, of the
node with `the maximal ID,' $Max(i)$. $\Delta_{tr}$
is a transmission time of the message $RootTime(j)$, sent from $j$
to $i$. Furthermore, we assume that $\Delta_{tr}$ is the fixed
transmission time for any $i\neq j$, and the message
$RootTime(j)$ is transmitted during that period of time;
Line 7: We let all other nodes, but $j$, know about
the recent updates $RootTime(i)$ and $Max(i)$.
Line 11: Finally, set the own clock.

The communication over the graph {\em CommGraph} is possible for every
node, since {\em CommGraph} is built as a concatenation in time of
$D$ identical copies of $\widetilde{G}$.

Let us now prove the correctness of Algorithm~\ref{alg:synchalgo}. With high
probability all $ID_i$ are different. There is a unique
node, let us call it $root = \max_{i \in [n]} ID_i$. We have to
show that all nodes in the network, after the synchronization
algorithm, have the same time, synchronized to the time of the
node $root$. Let us consider any node $i$ in the network. Since
the graph distance between the $root$ and $i$ is less than equal to
the diameter $D$, it follows that the entire synchronization
procedure will be done in $1+D+1$ steps, and all nodes will know the
time of the $root$. After Subroutine C (Set the Clock), all nodes will set their
own clocks, $OwnTime(i) = RootTime(i)$, all being equal to $OwnTime(root)$ with high probability. This proves the correctness of the Synchronization Algorithm, which is given in 
Fig.~2. 

\begin{figure}[htb]
\label{fig-commmunication}
\begin{minipage}[h]{0.55\linewidth}
\includegraphics[width=7cm]{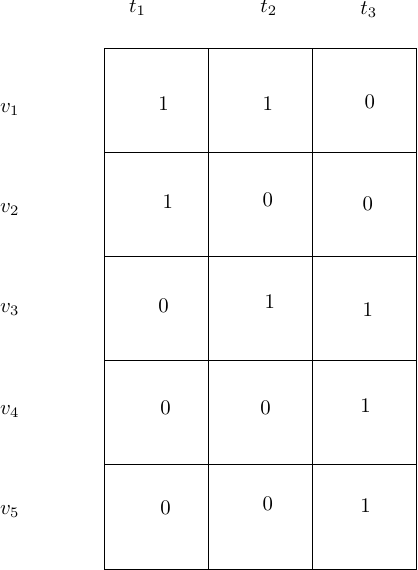}
\end{minipage}
\begin{minipage}[h]{0.40\linewidth}
\includegraphics[width=6cm]{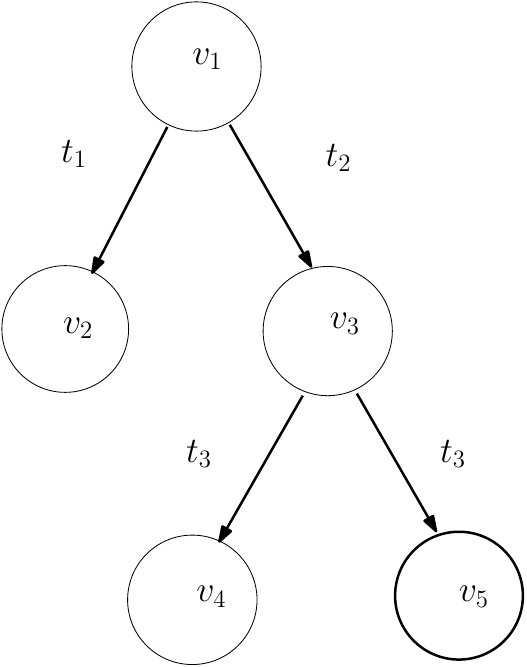}
\end{minipage}
\caption{At time $t_1$, nodes $v_1$ and $v_2$ can communicate; at time $t_2$, nodes $v_1$ and $v_3$ can communicate; at time $t_3$, nodes $v_3, v_4,$ and $v_5$ can communicate.}
\end{figure}

\section{Protocol That Does Not Need to Know the Number of Processors}

Suppose our processors know the offset $d$ but not the number of all processors in the system, that is, $n$.
The main observation here is that once we make a spanning tree of the graph, each node
can also compute the number of nodes in its spanning tree. Hence,
we can make an estimate of $n$ and then check to see if this estimate is
too big. Thus, until the right (within a constant factor) estimate
is reached, all nodes will reject the estimate and continue. Adjusting constants
appropriately, we can guarantee that an arbitrary constant fraction
of the processors will terminate with the right estimate of $n$
(within some fixed constant fraction). The algorithm for the estimation of $n$ is as
follows.

\begin{algorithm}[h!]
\caption{Estimation of $n$}
\label{alg:estimm}
\begin{algorithmic}[1]
\State
Set $i=0$.
\vspace*{1mm}
\State
Build a spanning tree
using the Main Algorithm (see Section~\ref{sec:main.algo}) for $n_i=d/2^i$ and
count the number of nodes in the tree. If the number of the nodes in the
tree is less than $n_i$, then set $i:=i+1$ and go to step 2.
\vspace*{1mm}
\State
Output $n_i$.
\vspace*{1mm}
\end{algorithmic}
\end{algorithm}

\begin{theorem}
Any constant fraction of the processors can synchronize
without knowing $n$, yet still use $O(d^{\frac{1-\beta}{2}}
\mbox{{\em {\em poly-log}}}(d))$ radio send/receive steps (with
probability of failure exponentially close to zero). The bound on the radio
use holds only for processors that synchronize.
\end{theorem}

\begin{proof}
We showed that the used power is $O\left(d^{\frac{1-\beta_i}{2}} \mbox{{\em poly-log}}(d)\right)$ for a particular number of processors $n_i = d^{\beta_i}$. Let us consider $n_i=d/2^i$. Since $\alpha_i = (1-\beta_i)/2$, it follows $d^{\alpha_i} = \sqrt{d/n_i}=2^{i/2}$. Let $i_{max} = \lceil \log
(d/n) \rceil $. Then the total power, used in the protocol that
does not know $m$, is
\begin{eqnarray}
\nonumber \sum_{i=0}^{i_{max}} O(d^{\alpha_i} \mbox{{\em {\em poly-log}}}(d)) &=&
\sum_{i=0}^{i_{max}} O(2^{i/2} \mbox{{\em {\em poly-log}}}(d))\\
\nonumber &=& \sum_{i=0}^{i_{max}} 2^{i/2} O (\mbox{{\em {\em poly-log}}}(d))\\
\nonumber  &=& O(2^{i_{max}/2}) O (\mbox{{\em {\em poly-log}}}(d)) \\
\nonumber &=& O(n^{(1-\beta)/2} \mbox{{\em {\em poly-log}}}(d) ),
\end{eqnarray}
since $2^{i_{max}/2} = O(\sqrt{d/n}) =  O(d^{\frac{1-\beta}{2}})$.
\end{proof}

\section{Our Model Can Handle Different Clock Speeds with Bounded Ratio}
\label{sec:clockspeeds}
In this section, we present the technical details that explain why our model is realistic even
if processors have somewhat different clock speeds.
For $n$ processors, let their clock speeds be $\{ \tau_1, \tau_2,
\dots, \tau_n\}$. Let $\tau_{\min}, \tau_{\max}$ be minimum, maximum
of the set $\{ \tau_1, \tau_2, \dots, \tau_n\}$, respectively. The
clock speeds are in general different, but the ratio
$\tau_{\max}/\tau_{\min} \leq c$ is bounded by some constant $c$,
and each processor knows that upper bound $c$. Let $\tau_{\textrm{trans}}$ be the lower bound on the time
necessary for the transmission, i.e., on the time necessary for communication and synchronization between two processors.
It is also assumed that  the lower bound on $\tau_{\textrm{trans}}$ is known to all processors. Now, knowing
$c$ and $\tau_{\textrm{trans}}$, each processor $i$ counts $k_i=2 c \tau_i \tau_{\textrm{trans}}$ clock
ticks as a single {\em time step} $s_i$. In other words, each processor enables the
condition necessary for the communication by the slowest processor. It follows that if two
processors $i$ and $j$ overlap for a period of time $ \geq
\min\{s_i, s_j\}/2$, then they can communicate.

For the purposes of analysis only, we assume that there is a global time axis, and time is mapped to the set
of non-negative real numbers. Note that there is no real global time, i.e., neither processors know nor need a real global time clock.
Let $s_{\max}:=\max \{s_1, s_2, \dots, s_n \}$. We now define a single unit of length $5 s_{\max}$ on the global time axis, which we call
a \textit{``time unit''}.
\begin{claim}
For every processor $i$ that works within a single global ``time unit''  there are at least three complete ``time steps'' that this processor's radio is awake.
\end{claim}

\begin{proof}
For the processor $i$ that starts working at some time $\zeta_i \in [0,s_i)$ within a single unit time $u$, the following is satisfied $\zeta_i + 3 \leq 5$, since we had previously defined the global time unit to be $5 s_{\max}$.
\end{proof}

\begin{lemma}
\label{claim:two-processors}
If two processors $i,j$ work within the same global time unit, then they can communicate and can synchronize.
\end{lemma}

\begin{proof}
Let us consider one time unit $u$ (with length of $5 s_{\max}$). Let $\zeta_i \in [0,s_i)$ be the time where the processor $i$ starts working within the unit $u$,
and analogously let $\zeta_j \in [0,s_j)$ be the time where the processor $j$ starts working within the unit $u$.
We argue that if they both happen to be awake in the same time unit, there is an overlap of time $\geq \min\{s_i, s_j\}/2$ when they both work and hence can communicate and can synchronize.

The processors $i,j$ certainly start working at times $\zeta_i \in [0,s_i)$, $\zeta_j \in [0,s_j)$, respectively, and then continue working over
the period $u$. Let $s_i \geq s_j$ (the other case is symmetric).
By Lemma~\ref{claim:two-processors} both processors work at least for three full-time steps. Then it
follows that there exist time instances $a>b$ within the unit $u$ such that: $i$
works over periods $[a-\tau_i, a]$ and $[a, a+ \tau_i]$; and $b \in [a-s_i, a]$ and $j$ works
for over periods $[b-s_j, b]$ and $[b, b+s_j]$. We pursue the analysis as follows.

If $a - b \geq s_j$ then $j$ is entirely covered by $i$.
Let us now analyze the case $a - b < s_j$.
Consider two time intervals $[a-s_i, a]$ and $[a, a+ s_i]$ when
$i$ works, as well as two intervals $[b-s_j, b]$ and $[b,
b+s_j]$ when $j$ works.
For $(a-b) \leq s_j/2$ then $b+s_j - a = s_j - (a-b) \geq s_j/2$,
otherwise $(a+s_i) - (b+s_j) = (s_i-s_j) + (a-b) > s_j/2$,
which completes the proof.
\end{proof}

\section{Conclusions and Follow-up Work}
In this paper, we have studied an important problem of power
consumption in radio networks and completely resolved the
deterministic case for two processors, showing matching upper and
lower bound. For multiple processors, we were able to show a
poly-logarithmic gap between our randomized protocol and our
deterministic lower bound. However, this is not completely
satisfactory. Our lower bound holds only for deterministic
protocols, while our upper bound in multi-processor case is
probabilistic (unlike the two-processor case, where our upper bound
is deterministic as well).

In the follow-up work to ours \cite{BDO10,BDO11}, the authors resolve this main
open problem posed by our work and show how to
achieve a deterministic upper bound of $O(d^{\frac{1-\beta}{2}})$, that  exactly matches
our deterministic lower bound of $\Omega(d^{\frac{1-\beta}{2}})$ for multiple processors and
answering the main open question left in our paper.

It is important to note that in radio communication conservation of power can be
achieved in two different ways. The first approach is to always broadcast
the signal with the same intensity (or to power down radios
completely in order to save energy);
it is the approach which we have explored in this work.
The second approach is the ability for a radio to
broadcast and receive signals at different intensity; the stronger
the signal the further it reaches. In the case where all processors
are at the same distance from each other, this is a non-issue (i.e.,
our single-hop networks, which is the main focus of this work). However, for
multi-hop networks the question of optimal power-consumption
strategies with varying signal strength is still completely open.

\end{document}